\documentclass[a4paper, conference]{IEEEtran}
\IEEEoverridecommandlockouts
\usepackage{cite}
\usepackage{amsmath,amssymb,amsfonts}
\usepackage{algorithmic}
\usepackage{graphicx}
\usepackage{textcomp}
\usepackage{xcolor}
\usepackage{multirow}
\def\BibTeX{{\rm B\kern-.05em{\sc i\kern-.025em b}\kern-.08em
    T\kern-.1667em\lower.7ex\hbox{E}\kern-.125emX}}
\begin{document}

\title{Exploiting Change Blindness for Video Coding: Perspectives from a Less Promising User Study \\
 }

\author
{\IEEEauthorblockN{Mitra Amiri}
\IEEEauthorblockA{\textit{ColorLab, dept. of Computer Science} \\
\textit{Norwegian University of Science and Technology (NTNU)}\\
Gjøvik, Norway \\
mitra.amiri@ntnu.no}
\and

\IEEEauthorblockN{Steven Le Moan}
\IEEEauthorblockA{\textit{ColorLab, dept. of Computer Science} \\
\textit{Norwegian University of Science and Technology (NTNU)}\\
Gjøvik, Norway \\
steven.lemoan@ntnu.no}
\and
\IEEEauthorblockN{Christian Herglotz}
\IEEEauthorblockA{\textit{Chair of Computer Engineering} \\
\textit{Brandenburgisch-Technische Universität Cottbus-Senftenberg}\\
Cottbus, Germany \\
christian.herglotz@b-tu.de}
 }

\maketitle

\begin{abstract}
What the human visual system can perceive is strongly limited by the capacity of our working memory and attention. Such limitations result in the human observer’s inability to perceive large-scale changes in a stimulus, a phenomenon known as change blindness. In this paper, we started with the premise that this phenomenon can be exploited in video coding, especially HDR-video compression where the bitrate is high. We designed an HDR-video encoding approach that relies on spatially and temporally varying quantization parameters within the framework of HEVC video encoding. In the absence of a reliable change blindness prediction model, to extract compression candidate regions (CCR) we used an existing saliency prediction algorithm. We explored different configurations and carried out a subjective study to test our hypothesis. While our methodology did not lead to significantly superior performance in terms of the ratio between perceived quality and bitrate, we were able to determine potential flaws in our methodology, such as the employed saliency model for CCR prediction (chosen for computational efficiency, but eventually not sufficiently accurate), as well as a very strong subjective bias due to observers priming themselves early on in the experiment about the type of artifacts they should look for, thus creating a scenario with little ecological validity.       
\end{abstract}


\section{Introduction}
Although one might think that we see the surrounding world in great detail, it has been proved the human visual system (HVS) limits what we see due to limitations in our physiology as well as our working memory \cite{b1}. As a result, human observers tend to miss rather big and visible changes in their visual field. This phenomenon is usually referred to as change blindness (CB) \cite{b2, b3}. Change blindness is dependent on individual parameters such as age, alertness, and even familiarity of the individual with the scene as well as the scene complexity \cite{b23}. The observer's failure to pick up information from some regions of a scene (Image or video) can be translated into the presence of irrelevancies. These irrelevancies are what lossy compressions take advantage of. 

Among the common media sources, raw videos have some of the largest files. During the last decade, the growth of streaming services \cite{b4} has led to increased demand for efficient codecs with high video-quality output. In addition, the growth in the consumer's interest in high dynamic range (HDR) technology has pushed streaming services to provide HDR video streams \cite{b5}. However, HDR video files are potentially larger than standard dynamic range (SDR) videos and would require heavier compressions in order to be streamed. The present compression methods such as advanced video coding (AVC)\cite{b6} and high-efficiency video coding (HEVC)\cite{b7} which are extensively used, use block-based motion compensation and content-adaptive entropy coding to encode the video. However, in none of the cases, the quantization parameter (QP) is spatially or temporally alternated based on saliency and attention.    
One of the earliest works done in order to include attention in video compression was done by Li et al. \cite{b8} where they suggested the modification of the bit allocation strategy in AVC encoding, based on the output of a saliency detection algorithm. In another work, Hadizadeh et al. \cite{b9} introduced the region-of-interest (ROI)-based method that uses the Itti-Koch-Niebur saliency model \cite{b10} to incorporate the saliency into the direct cosine transform (DCT) domain in the AVC encoding. In a similar work, Barua et al. \cite{b11} suggested using saliency maps to create high-fidelity regions incorporated into JPEG and MPEG-4 encoding for low-bitrate surveillance images and videos. However, despite the effectiveness of the discussed approaches, none of them has tried to incorporate CB. Saliency as an output of attention, although relevant to CB is not the only parameter affecting the observer's perception of the visual stimuli. Several studies have demonstrated the contribution of the limitations of visual working memory to CB \cite{b20, b21}. In a recent work, Le Moan and Farup \cite{b12} took a further step and suggested a specifically defined change blindness map and used it to reduce the redundancies in regions with a high chance of CB occurrence. To our understanding, CB-based encoding has not yet been used in video compression.

In addition to the effect of saliency on CB, the graduality of the change is also effective in the change detection ability of HVS. The effect of graduality is most visible in the case of the "spot the difference" task. It has been shown that if the two images are shown in a fast-flickering manner, the observers can spot the difference quickly while if the frequency of the flickering is reduced, the observers struggle with finding the difference between the two images \cite{b3}. In a recent study, Alizadeh et al. \cite{b13} demonstrated that the human observer tends to fail following the gradual change. 

In this work, we would initially introduce our suggested novel algorithm of spatiotemporal encoding. We explore the effect of several parameters such as the introduction of CCR, the QP schedule curve, and the duration of the QP variation utilizing a subjective video quality assessment experiment. Afterward, we would focus specifically on the takeaways from our unfavorable subjective experiments.  
 
\section{Materials and methods}

\subsection{Video Dataset}

\begin{figure}[t]
\centerline{\includegraphics [width=0.5\textwidth]{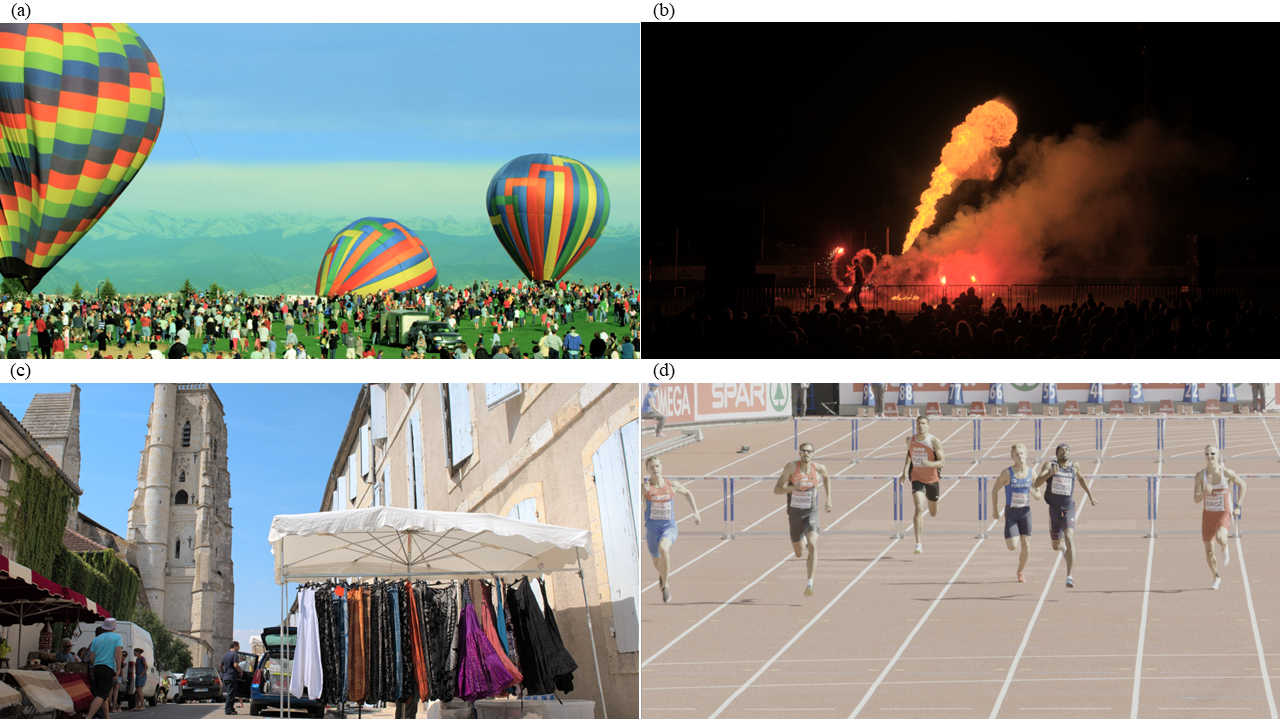}}
\caption{The four video sources used in this study: a) Balloon Festival (9.96s), b) Fire eaters (8.3s), c) Market (16.62s), and d) Runners (20.8s).}
\centering
\label{fig-1}
\end{figure}

The videos used in this study were taken from the JEVT dataset \cite{b14}. A total of four raw 10-bit videos in YUV format were selected as the source video sequence. The videos were selected in a way that they would cover both still and highly dynamic scenarios, crowded and uncrowded scenes, and the presence and absence of dominant human faces. All videos had a resolution of 1920×1080 pixels and a frame rate of 24 frames per second (fps). The duration of the video sources varied from 8.3s (199 frames) to 20.8s (499 frames). Figure ~\ref{fig-1} shows the four source videos employed for this study. 

\subsection{CCR Prediction Algorithm}

\begin{figure}[t]
\centerline{\includegraphics [width=0.5\textwidth]{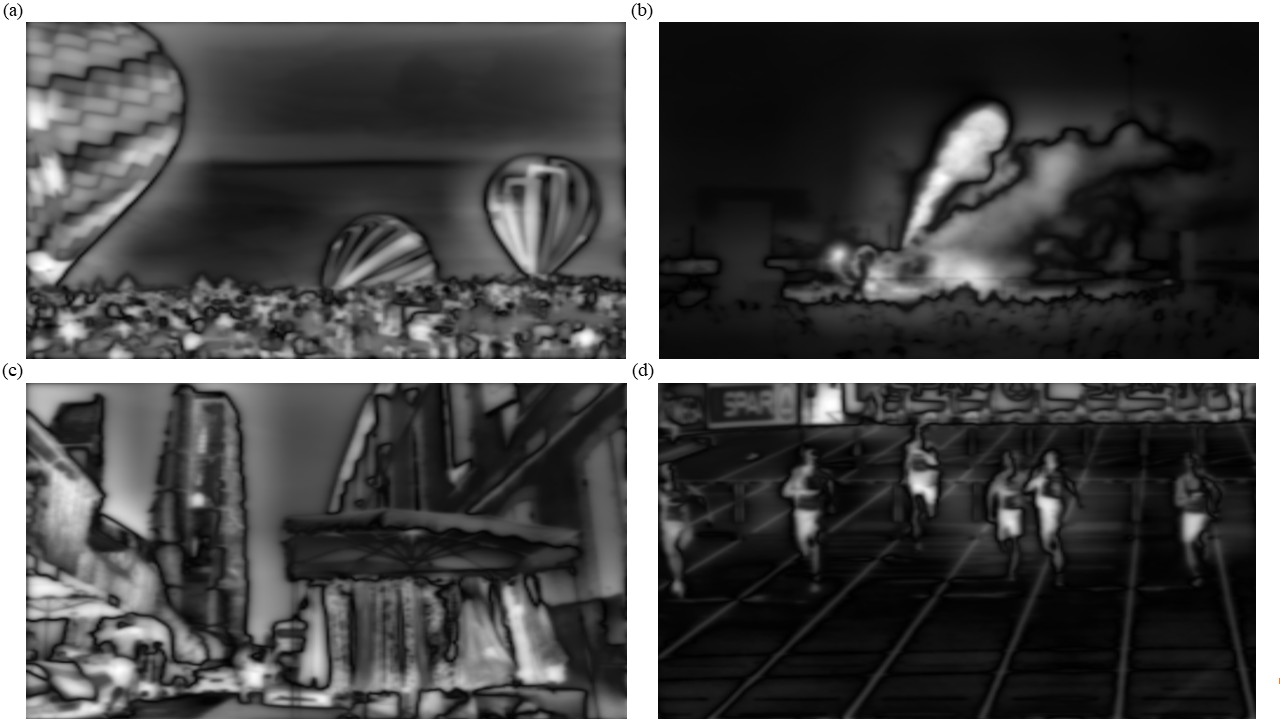}}
\caption{Examples of the saliency maps used to define the CCRs: a) Balloon festival, b) Fire eaters, c) Market, and d) Runners.}
\centering
\label{fig-3}
\end{figure}

To enable the spatially  \(QP\)-varying coding, it was necessary to predict the regions where CB was most probable to happen. These regions are referred to as Compression Candidate Regions (CCR). The prediction of CCR was carried out per frame and followed the SDSP algorithm suggested by Zhang et al. \cite{b14}. The algorithm relies on three types of priors, a frequency prior, a color prior, and a location prior. The frequency prior used in the SDSP algorithm is based on a research output of Achanta et al. \cite{b15}. The algorithm also counts for center bias which is simulated using a Gaussian map since according to the findings of Judd et al., \cite{b16}, human observers tend to pay more attention to the center of an image. Overall, the Saliency Detection by combining Simple Priors (SDSP) algorithm is computationally light and quite fast which makes it suitable for real-time encoding, yet as it can be seen, it is based on low-level visual features and does not take the semantic information of the scene into account. Additionally, SDSP does not account for the saliency of the moving objects and their tradeoff with center bias. Meanwhile, in the case of the chosen videos, in many instances, mobile attractive objects are present at the edges of the scene. Consequently, due to SDSP’s calculational speed, it was decided to proceed with this algorithm while excluding the center bias to partially account for the shortcomings of the saliency prediction algorithm. Figure ~\ref{fig-3} shows a sample of the saliency maps for each source video sequence. The complementary of these saliency maps were used to achieve the CCRs.   

\subsection{Video Encoding Tool}
The videos were encoded using the Kvazaar open-source HEVC encoder \cite{b17}. Although the HEVC is designed to use the P-frames for inter-frame prediction, to minimize the observed effect of the frame-encoding paradigm in the suggested method, the encoding parameters were set in a way that during the entire sequence, only one I-frame (Intra-frame) would exist (Table ~\ref{tab-1}). As a result, all videos were encoded by starting with an I-frame followed by bidirectional frames (B-frames) throughout the video sequence and no prediction frames (P-frame) were present. The spatiotemporal varying QPs were included using the (\texttt{--}roi) tag and a text file including the quantization parameter difference (\(\Delta QP\)) values which will be described in more detail in the following section. Finally, to reduce the encoding complexity, the ultrafast preset was used.   

\begin{table}[htbp]
\centering

\caption{The common parameters used in the Kvazaar encoder}
\begin{tabular}{l l l}
\hline
\textbf{Parameter} & \textbf{Tag} & \textbf{value} \\
\hline
Frequency of I-frame Repeat & \texttt{--}period & 0 \\
Length of the Group of Pictures (GoP) & \texttt{--}-gop & 0 \\
Video Resolution & \texttt{--}input-res & 1920$\times$1080   \\
Encoding Preset & \texttt{--}preset & Ultrafast \\
Bit Depth  & \texttt{--}input-bitdepth & 10 \\
Input Frame rate & \texttt{--}input-fps & 24 \\
Quantization parameter & \texttt{--}qp & 22 \\
\hline

\end{tabular}
\label{tab-1}\label{tab-1}
\end{table}

\subsection{Stimuli preparation} \label{SP}
To introduce the CCR as well as the temporal gradual change in the HEVC encoding algorithm, a map of \(\Delta QP\) for each frame was created. In the spatial domain, the \(\Delta QP\) which will be referred to as the RoI map was generated by resizing the CCR map to the desired number of blocks in each dimension using cubic interpolation. Afterward, the RoI map was multiplied by the \(\Delta QP\) values designated for each frame to create the spatially varying \(\Delta QP\) map for each frame. Finally, the \(\Delta QP\) map for all the frames was saved in a text file that would be used under the --roi tag. In our study, two scenarios were implemented. In the first scenario, the RoI map consisted of a single block, encoding the importance of the frame, and no spatial CCR was encoded. In the second scenario, each frame was divided using the block size (BS) of 10 (10 by 10 blocks). 

\begin{figure}[htbp]
\centerline{\includegraphics [width=0.5\textwidth]{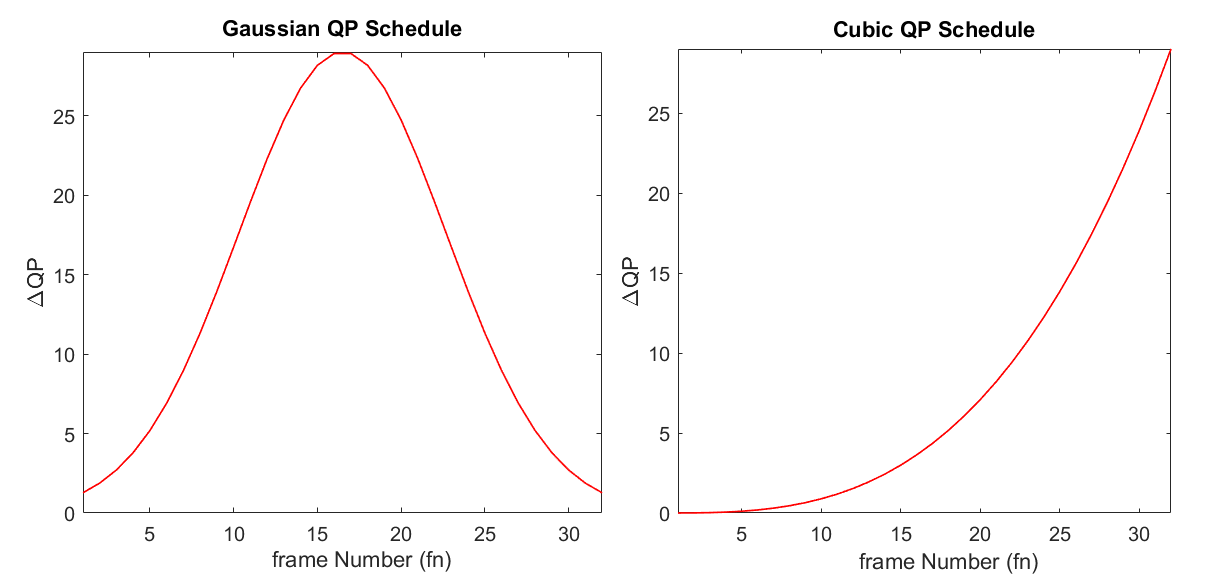}}
\caption{The example of the employed \(\Delta QP\) schedule over the frame period of 32 frames ranging between 0 to 29.}
\centering
\label{fig-4}
\end{figure}

On the other hand, to exploit the blindness of human observers to the temporal changes, a gradual \(\Delta QP\) schedule was employed that defined the designated \(\Delta QP\) value for each frame throughout the video sequence. Figure ~\ref{fig-4} shows the two main \(\Delta QP\) schedules used in our study.

The range of the frame numbers (nf) within which the gradual change happens varied between 16 frames, 32 frames, and the full video length. As for the \(\Delta QP\) range, the higher limit was dictated by Kvazaar and the baseline QP. In the case of the frame numbers lower than the full video length, the schedule would repeat throughout the video sequence. As a result, to prevent the sharp descent of QP at the end of the cycle (typical in cubic schedule), the cubic schedule was only used for the full video-length schedules. 

Based on Kvazaar’s implementation, the highest possible QP while encoding using Kvazaar is 51, and since the baseline QP was set to 22 (Table 1) the upper boundary of the \(\Delta QP\) range was set to 29. In order to push the video compression to an extent in which the compression artifacts are more visible, the lower boundary of the \(\Delta QP\) range was set to 15. 

The combination of two frame numbers with gaussian schedule and the two schedules for the full video length frame number, with the two RoI scenarios led to an overall 8 configurations.  In the next step, the size of each encoded video was used as a target in an optimization loop through which a conventional HEVC-encoded video with constant QP was generated. Since the QP values could only be integers and the video size was not a continuous function of QP, the closeness of the two video sizes was ensured by introducing the constraint by which the constant QP videos could not be more than 5\% smaller than the target video size. As a result, the final number of stimuli per image source was 16 videos and the overall number of stimuli was 64.

\subsection{Experiment design}
To evaluate the general quality of the encoded videos and compare the quality of the saptio-temporally varying encoded videos with the conventional HEVC encoded videos, a psychophysical experiment was carried out. The stimuli included the 64 encoded videos with varying configurations, the four source videos along with six randomly chosen videos to account for intra-observer variability. In order to have the distance between the quality of different stimuli, the single-stimulus hidden-reference category judgment scheme was chosen \cite{b18}. According to the International Telecommunication Union (ITU) recommendations \cite{b18}, 5 absolute levels were selected ranging from “Bad” to “Excellent”.

\begin{figure*}[t]
\centerline{\includegraphics [width=\textwidth]{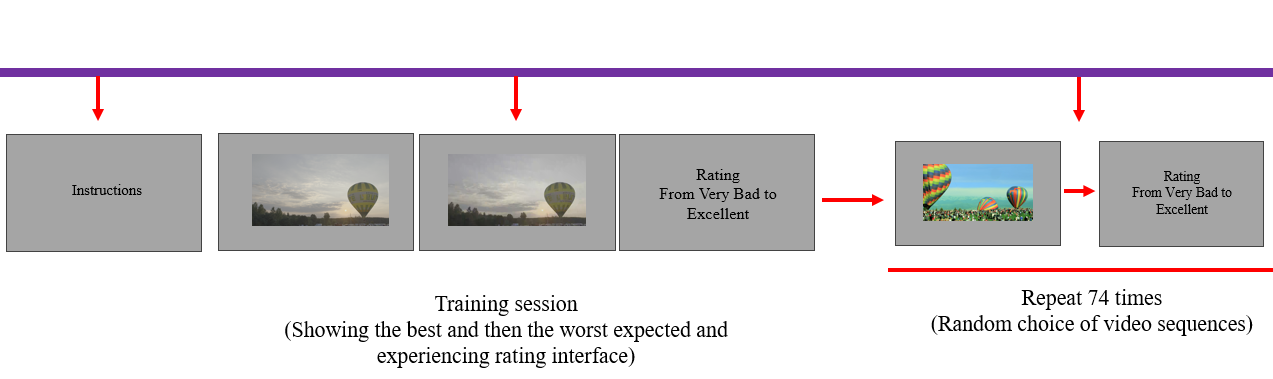}}
\caption{The scheme of the experiment. The experiment starts with a training session for the observers to get exposed to the best and the worst quality cases along with the rating interface, followed by the actual experiment which was run for 74 stimuli. }
\centering
\label{fig-5}
\end{figure*}

 Figure ~\ref{fig-5} presents the scheme of the experiment. As it is portrayed, the experiment was carried out in two phases. Before the training phase, the observers were presented with the general instructions for the experiment. In the next step, the observers went through the training phase where they were presented with an example of an excellent and a bad video sequence. The video source used for this step was the "SunRise" video, taken from the JVET HDR dataset \cite{b19}. The observers also got to experience the rating interface during the training phase.

After finishing the training section, the observers moved on to the main experiment where they were presented with the randomized stimuli video sequences one at a time and were asked to rate the videos between "Very Bad" to "Excellent". The participants could observe each video sequence once and they were given unlimited time to rate the videos. After finishing the experiment, the observers were interviewed about the element they mainly focused on while observing each video source. 

\subsection{Experiment setup}
In the experimental set up a 31.1-inch EIZO HDR CG3146 (4096$\times$2160) was used which was calibrated using its built-in camera to match DCI-P3. The observers' seat was set at 1m distance from the display and a desk was put between the observer and the display to make sure that the observer did not approach the display at any time during the experiment. The experiment session was designed using MATLAB (2022b). The order of the video display was randomized and the user rating interface was written in MATLAB yet, the playback was carried out using ffplay. Using ffplay it was ensured that the video sequence is displayed in the original resolution. Since the display's resolution was higher than the resolution of the video sequence, the video sequence was shown over a neutral gray background.


\subsection{Observers}
In general, 16 observers (8 male and 8 female) participated in the experiment. All the observers were students or members of the IDI department at NTNU (Gjøvik) and their range of expertise varied from familiar to highly experienced in the field of image and video processing. The observer's age varied from 22 to 36 years old and all had normal or corrected to normal vision.  

\subsection{Variability studies}

In this study, the agreement percentage was employed as a measure of inter-observer variability to evaluate the variability of opinions among the observers. Agreement percentage can be expressed as the ratio of cases in which the observer's opinion score fell within the range of \(\mu_{n-1}\pm\sigma_{n-1}\) (where \(\mu_{n-1}\)is the mean opinion score of all observers other than the studied observer and \(\sigma_{n-1}\)is the standard deviation) to all the observed cases. 
On the other hand, the consistency of observers in their opinions was calculated using the mean of the absolute opinion score distance \((\overline{|\Delta S|} \)). Equation ~\ref{eq} presents the mean of absolute opinion score distance where $m$ is the number of repeated stimuli and \(S_i\) is the opinion score for the \(i^{th}\) sample, Orig stands for original, and RP stands for repeated. 

\begin{equation}
\overline{|\Delta{S}|}= \sum{|S_{i,Orig}-S_{i,RP}|}/m\label{eq}
\end{equation}

\section{Results and discussion}
\subsection{Observer Variabilities}
\begin{figure}[htbp]
\centering
\includegraphics[width=0.9\linewidth]{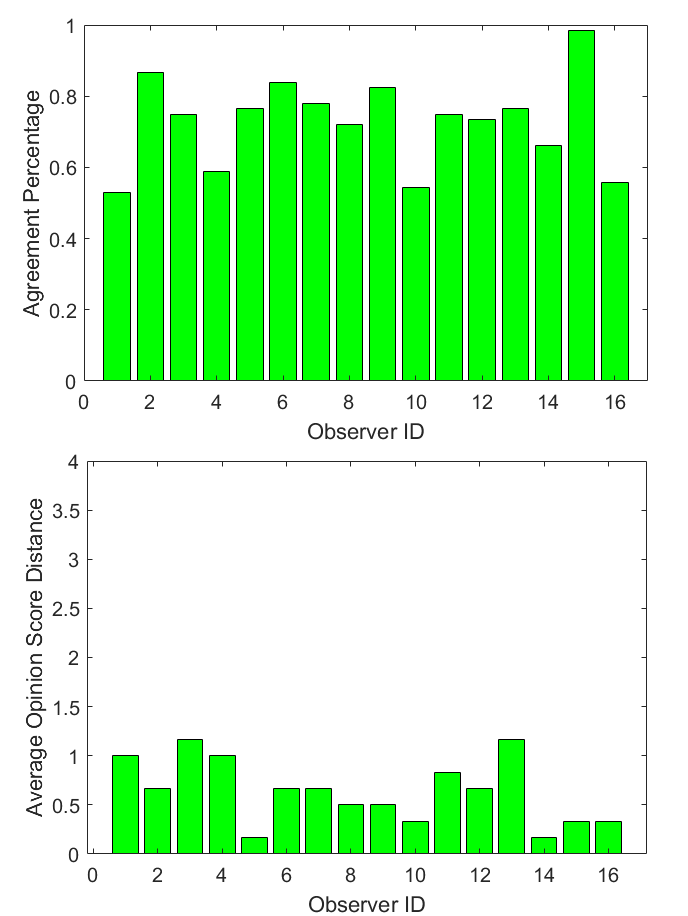}

\caption{ Observer Variability results: Top) Inter-observer variability expressed in terms of agreement percentage, Bottom) Intra-observer variability expressed in terms of absolute opinion distance.}
\label{fig-8}
\end{figure}

Figure ~\ref{fig-8}  shows the inter- (top) and intra-observer variability (bottom) among the 16 observers. As it can be seen, four observers (1,4,10 and 16) show high inter-observer variability in comparison with other observers. However, their agreement percentage is more than 50\%, and looking at the intra-observer variability of these observers, none of them showed a mean opinion distance of higher than 2 which makes them reliable enough to use the data collected from them in further evaluations.  

\begin{table} [b]
\centering
\caption{The absolute opinion score difference in the case of each repeated stimulus.}
\begin{tabular}{l l l l}
\hline
\textbf{Parameter} & \textbf{Mean} & \textbf{Standard deviation} & \textbf{Range} \\
\hline
BalloonFestival-nf-16-BS-10-G & 0.5000 & 0.6325 & [0-2] \\
BalloonFestival-nf-239-BS-10-P3 & 0.6250 & 0.7188 & [0-2] \\
FireEaters-nf-32-C & 0.6250 & 0.7188 & [0-2] \\
Market-nf-32-BS-1-G & 0.5000 & 0.6325 & [0-2] \\
Market-nf-399-BS-10-G & 0.8750 & 0.7188 & [0-2] \\
Runners-nf-16-BS-10-C & 0.6875 & 0.7042 & [0-2] \\
\hline

\end{tabular}
\label{table_for_fig7}
\end{table}


Looking at the absolute opinion differences per each sample (Table. ~\ref{table_for_fig7}), it can be seen that none of the chosen samples were highly confusing (more than 2 score distances) for all the observers. As a result, the present intra-observer variability can be related to the observer's reliability.

\subsection{Source Video Quality Ratings}
Table  ~\ref{tab-2}, portrays the MOS values for the source video sequences along with the size of the source videos. All the source videos approximately had the same bitrate of 1.2 Gbps. It can also be seen that all the source video sequences have been rated highly. Furthermore, no statistically significant quality difference was visible between the source videos. 

\begin{table} [h]
\centering

\caption{The MOS values and the file sizes of the source videos}
\begin{tabular}{lcccc}
\hline
\textbf{Sources}& \textbf{Balloon Festival} & \textbf{Fire Eaters} & \textbf{Market} &\textbf{Runners} \\
\hline
\textbf{MOS} & 4.625 & 4.687 & 4.937 & 4.500 \\
\textbf{SD} & 0.500 & 0.793 & 0.250 & 0.894 \\
\textbf{Size (Mb)} & 1493  & 1244 & 2883 & 3110 \\
\hline
\end{tabular}
\label{tab-2}
\end{table}



\begin{table*}[ht] 
\centering
\caption{Bitrate of each video, the bitrates highlighted in red are the lowest per source video. Note: "G" stands for Gaussian, "FL" stands for full length, 'P3' stands for Cubic, and "C-QP" represents the video encoded with constant QP and matching size.}
\begin{tabular}{| l | l  l | l l | l l | l l |}
\hline
\multirow{2}{*}{\textbf{Source Video}} & \multicolumn{8}{l|}{\textbf{Bitrate (Mbps)}} \\
\cline{2-9}
 & \multicolumn{2}{l|}{nf=16/G    C-QP} & \multicolumn{2}{l|}{nf=32/G   C-QP} & \multicolumn{2}{l|}{nf=FL/G   C-QP} & \multicolumn{2}{l|}{nf=FL/P3    C-QP} \\
\hline
{Balloon-No CCR} & 3.4417 & 3.0699 & 1.8011 & \textcolor{red}{\textbf{1.6478}} & 3.3116 & 3.0699 & 2.5277 & 2.2496 \\
{Balloon-with CCR} & 5.6153 & 5.6047 & 3.7908 & 3.0699 & 5.1243 & 4.2431 & 4.6629 & 4.2431 \\
{FireEaters-No CCR} & 0.6965 & 0.5696 & 0.3688 & \textcolor{red}{\textbf{0.3551}} & 0.6778 & 0.5696 & 0.5126 & 0.4489 \\
{FireEaters-with CCR} & 1.8064 & 1.4832 & 1.4928 & 1.4832 & 1.7894 & 1.4832 & 1.6297 & 1.4832 \\
{Market-No CCR} & 4.3598 & 4.1944 & 4.2153 & 4.1944 & 2.0041 & \textcolor{red}{\textbf{1.6931}} & 3.4344 & 3.3762 \\
{Market-With CCR} & 6.9404 & 6.4294 & 6.8057 & 6.4294 & 4.3588 & 4.1944 & 5.9002 & 5.1758 \\
{Runners- No CCR} & 1.0263 & 1.0233 & 1.0101 & 0.8950 & 0.5178 & \textcolor{red}{\textbf{0.4683}} & 0.7288 & 0.5984 \\
{Runners-With CCR} & 2.0051 & 1.9030 & 1.9805 & 1.9030 & 1.3772 & 1.1893 & 1.6727 & 1.3818 \\
\hline
\end{tabular}
\label{tab-3}
\end{table*}

\begin{figure*}[ht]
\includegraphics [width=1\textwidth]{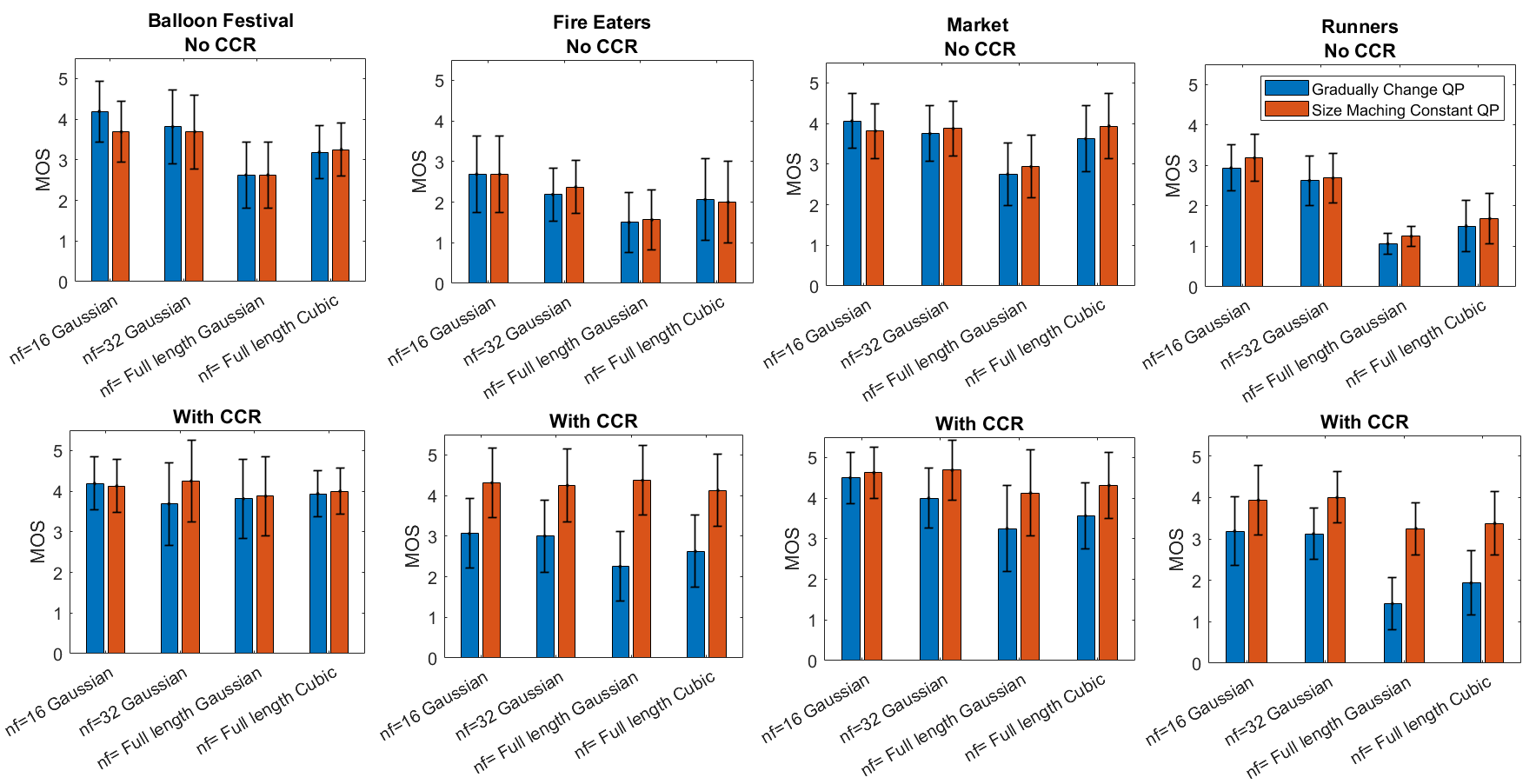}
\caption{Comparison between the spatiotemporally encoded videos and their best size-matching HEVC videos.}
\centering
\label{fig-11}
\end{figure*}

\subsection{Discussion}
As described in section ~\ref{SP}, to answer the question of whether the proposed encoding algorithm outperforms the current HEVC encoding in terms of video quality; for each spatiotemporally encoded video, a conventionally HEVC encoded video (with constant QP) with the same size was generated. Table  ~\ref{tab-3} lists the bitrate of the spatiotemporally encoded video along with its best size matching HEVC encoded video for each source video and each configuration scenario. As it can be seen the constant-QP size matched videos have a bit rate of less than 20\% lower than their paired gradually changing-QP encoded videos. 

Figure ~\ref{fig-11} summarizes the quality MOS. In this figure, each column presents the cases for one source video sequence, each row represents one spatial encoding scenario, and within each diagram, different \(\Delta QP\) schedules are presented. Looking at the graphs in Figure ~\ref{fig-11}, it can be seen that in none of the cases, the spatiotemporal encoding has led to significantly higher quality than the size-matched HEVC-encoded videos. Even in the case of the 16-frame Gaussian gradual change, the difference was reported as not significant based on the t-Student test (P =0.0926).

As it can be seen, either the qualities are rated closely or the constant-QP video is significantly better is content-dependent. However, in the case of temporal-only encoding (no CCR), no significant difference between the quality of the proposed encoding and the HEVC encoding was observed. While in the presence of CCRs, significantly lower quality was observed in the case of the spatially encoded videos with dynamic scenes (FireEaters and Runners). Based on the observed difference, it can be suggested that the saliency prediction algorithm fails to include the saliency raised from the movement of the objects. This conclusion is aligned with the points noticed in the interviews as the observers could obviously see the compression artifacts at the edges of moving objects such as the runners or the flames. Looking at the CCR maps, the presence of dark edges right next to highly salient regions would lead to strong compression artifacts at a very close distance to the salient regions. As a result, there is a high chance that the presence of the compression artifacts would capture the observers’ attention. The mentioned observations suggest that to successfully exploit change blindness using our proposed method the employed saliency prediction algorithm should predict the human observer's saliency very closely and account for the temporally-raised saliency.  

Nevertheless, it can be seen that both in the presence and absence of spatial \(\Delta QP\) variation, in some video sources, the increase in the frame number has led to a significant decrease in the average perceived quality (between 16 and full video length frames). This can be due to the fact that when the gradual change occurs over a longer time frame, a bigger number of frames with low quality are consecutively shown to the observers, therefore, the observer has a longer time to recognize the artifacts and associate it with lower quality. Furthermore, a higher average perceived quality is observed in the case of the cubic schedule in comparison to the Gaussian schedule across the full video length while due to the large standard deviation, the difference is not significant. This observation can be explained by the fact that in the case of the Gaussian schedule, the highly compressed frame occurs in the middle of the video sequence leaving the observers a longer time to incorporate their observation with lower quality. Based on these observations, the frame number employed in the suggested algorithm is required to be optimized based on the \(\Delta QP\) temporal schedule and in a way that the gradually appearing compression artifacts would escape observers' attention.  

By the comparison between the MOS values and the video sizes for the with and without CCR scenarios, it can be observed that the presence of CCR in encoding has led to higher bitrates. As a result, while the quality of the spatiotemporally encoded videos has not significantly changed, the size-matching HEVC files have quite larger sizes and therefore higher quality. Furthermore, the observers frequently reported paying attention to parts of the scene that the CCR map generally marks as less important. For instance, in the case of the Market, the observers paid a lot of attention to the wall and the tower, and the umbrella while the prediction model does not include any of these as highly salient regions. In another example, the observers frequently reported the runners' faces and the text on their shirts as important regions while the model notes the runner's shorts as the most salient regions. These observations suggest that not only does the SDSP algorithm fail in the case of dynamic scenes, but it also fails to properly mimic human attention even in less dynamic scenes such as the market. However, it must also be noted that all of the observers were to some extent experts in the field of image and video quality, and their attitude toward the task was affected by the bias caused by their expertise \cite{b22}. Based on the post-experiment interviews they were actively searching for regions where the artifacts were expected. It is known that CB is affected by attention ~\cite{b20} and if the attention of the observer due to the task architecture or due to the expertise, is drawn to the locations where change is present, exploitation if CB becomes impractical. It is worth mentioning that although the common quality measurement methods are well-established, they are designed to measure the perceived quality in scenarios of low-level visual masking where the observer's attention does not play a major part in the perceived quality. Our observations suggest that the evaluation of perceived video quality in scenarios where high-level visual masking is present would require a new methodology where instructions, multi-sensory stimuli or distractors prevent the effect of expertise and task-allocated attention.

\section{Conclusion}

Based on the principles of change blindness, a novel spatiotemporal video encoding approach for HDR and SDR video content within the framework of HEVC encoding was proposed. The subjective quality of the encoded videos was then compared to the quality of HEVC videos of the same size to determine the effectiveness of the proposed method. However, no significant enhancement in the subjective quality of the spatiotemporally encoded video was observed and even in some cases, the HEVC encoded video significantly outperformed the suggested encoding method. The results illustrated that the use of a proper saliency prediction algorithm for the prediction of the compression candidate region plays a key role in the proposed algorithm. It was observed that the saliency prediction should include both temporal and semantic saliency in order to be successful in exploiting change blindness. Additionally, it was observed that if the gradual change is carried out over a shorter period, the observers perceive less drop in the quality.         
Although the initial results did not show significant enhancement in the spatiotemporally encoded video, the current work can be used as a ground to explore better and more semantic CCR prediction models. Additionally, in this work, the effect of block size and its correlation with the CCR map accuracy was not studied.  Exploring these parameters can potentially lead to the enhancement of the proposed algorithm. Additionally, it was shown that the use of experts in the subjective study could considerably affect the results when exploring change blindness applications due to the alteration of their attention allocation based on their expertise-based bias. Finally, it was observed that the common quality measurement methodology is less suitable for evaluation of attention-dependent perceived quality.

\end{document}